\documentclass[pra,preprint,superscriptaddress]{revtex4-1}
\usepackage{amsmath}
\usepackage{theorem}
\theorembodyfont{\normalfont}
\newtheorem{theorem}{Theorem}
\newtheorem{definition}{Definition}

\begin{document}

\title{Sampling of globally depolarized random quantum circuit}
\begin{flushright}
YITP-19-90
\end{flushright}
\author{Tomoyuki Morimae}
\email{tomoyuki.morimae@yukawa.kyoto-u.ac.jp}
\affiliation{Yukawa Institute for Theoretical Physics,
Kyoto University, Kitashirakawa Oiwakecho, Sakyoku, Kyoto
606-8502, Japan}
\affiliation{JST, PRESTO, 4-1-8 Honcho, Kawaguchi, Saitama,
332-0012, Japan}
\author{Yuki Takeuchi}
\email{yuki.takeuchi.yt@hco.ntt.co.jp}
\affiliation{NTT Communication Science Laboratories,
NTT Corporation, 3-1 Morinosato Wakamiya,
Atsugi, Kanagawa 243-0198, Japan}
\author{Seiichiro Tani}
\email{seiichiro.tani.cs@hco.ntt.co.jp}
\affiliation{NTT Communication Science Laboratories,
NTT Corporation, 3-1 Morinosato Wakamiya,
Atsugi, Kanagawa 243-0198, Japan}

\date{\today}
\begin{abstract}
The recent 
paper [F. Arute et al. Nature {\bf 574}, 505 (2019)]
considered exact classical sampling of 
the output probability distribution of
the globally depolarized
random quantum circuit. 
In this paper, we show three results.
First, we consider the case
when the fidelity $F$ is constant. 
We show that if the distribution is
classically sampled in polynomial time within a constant
multiplicative error, then
${\rm BQP}\subseteq{\rm SBP}$,
which means that BQP is in the second level of the
polynomial-time hierarchy.
We next show that for any $F\le1/2$,
the distribution is classically trivially sampled by the uniform
distribution 
within the multiplicative error $F2^{n+2}$,
where $n$ is the number of qubits.
We finally show that for any $F$,
the distribution is classically trivially sampled by the uniform
distribution 
within the additive error $2F$.
These last two results show that if we consider
realistic cases, both $F\sim2^{-m}$ and $m\gg n$, 
or at least $F\sim2^{-m}$,
where $m$ is the number of gates,
quantum supremacy does not exist
for approximate sampling even with the exponentially-small errors.
We also argue that if $F\sim2^{-m}$ and $m\gg n$,
the standard approach will not work
to show quantum supremacy even for
exact sampling.
\end{abstract}

\maketitle

\section{Introduction}
Several sub-universal quantum computing
models are shown to be hard to classically simulate.
For example, output probability distributions of
the depth-four model~\cite{TD},
the Boson Sampling model~\cite{BS}, 
the IQP model~\cite{IQP1,IQP2},
the one-clean qubit 
model~\cite{KL,MFF,M,Kobayashi,KobayashiICALP},
and the random circuit model~\cite{random1,random2}
cannot be classically sampled in polynomial time unless some conjectures
in classical complexity theory (such as the infiniteness of the polynomial-time
hierarchy) are refuted.
Impossibilities of exponential-time classical simulations of sub-universal
quantum computing models
have also been shown recently based on classical fine-grained complexity
theory~\cite{Dalzell,DalzellPhD,Huang,Huang2,MorimaeTamaki}.

Let $p_z$ be the probability that an $n$-qubit ideal
random quantum circuit outputs the $n$-bit string $z\in\{0,1\}^n$.
Ref.~\cite{Google}
considered the globally depolarized version 
where 
the probability $p_z'$ that the output is
$z\in\{0,1\}^n$ is written as
\begin{eqnarray*}
p_z'=Fp_z+\frac{1-F}{2^n},
\end{eqnarray*}
where $0< F<1$ is the fidelity. 

In this paper, we show the following three results:

\begin{theorem}
\label{theorem:result}
Assume that $F$ is constant. Then,
if the probability distribution $\{p_z'\}_{z\in\{0,1\}^n}$
is sampled in classical $poly(n)$ time within a constant multiplicative
error $\epsilon<1$,
then ${\rm BQP}\subseteq{\rm SBP}$.
\end{theorem}

\begin{theorem}
\label{theorem:result2}
For any $F\le\frac{1}{2}$,
$\{p_z'\}_{z\in\{0,1\}^n}$ is classically sampled
by the uniform distribution within the
multiplicative error $F2^{n+2}$.
\end{theorem}

\begin{theorem}
\label{theorem:result3}
For any $F$,
$\{p_z'\}_{z\in\{0,1\}^n}$ is classically sampled
by the uniform distribution within the additive error $2F$.
\end{theorem}

Proofs are given in the later sections. 
In the rest of this section, we provide several remarks.

First, the class SBP~\cite{SBP} is defined as follows.

\begin{definition}
A language $L$ is in SBP if and only if there exist a polynomial
$s$
and a classical polynomial-time probabilistic algorithm
such that if
$x\in L$ then $p_{acc}\ge 2^{-s(|x|)}$, and
if $x\notin L$ then $p_{acc}\le 2^{-s(|x|)-1}$. 
Here, $p_{acc}$ is the acceptance probability.
\end{definition}

Note that the class SBP remains unchanged even when the
two thresholds, $2^{-s(|x|)}$ and $2^{-s(|x|)-1}$, are replaced with
$\alpha 2^{-s(|x|)}$ and $\beta2^{-s(|x|)}$, respectively, for any 
constants
$\alpha$ and $\beta$ satisfying $0\le\beta<\alpha\le1$.
It is known that SBP is in AM, and therefore ${\rm BQP}\subseteq{\rm SBP}$
means that BQP is in the second level of the polynomial-time hierarchy.
The containment of BQP in the polynomial-time hierarchy is not
believed. (For example, there is an oracle separation~\cite{Raz}.)

Second,
note that 
quantum supremacy for the globally depolarized circuits 
was previously studied
in Ref.~\cite{nonclean}
for the one-clean qubit model.

Third, Theorem~\ref{theorem:result} holds for a broader class
of quantum circuits than the 
globally depolarized random circuit.
In particular, we can replace our random gate application
with the coherent one.
In this paper, however, we concentrate on the 
globally depolarized random circuit
for the simplicity.
Theorem~\ref{theorem:result2} and
Theorem~\ref{theorem:result3} hold for the output probability
distribution $\{p_z\}_z$ of any quantum circuit.

Finally, 
in Ref.~\cite{Google}, it was claimed that if the exact polynomial-time
classical sampling
of $\{p_z'\}_z$ is possible, then estimating 
$|\langle0^n|U|0^n\rangle|^2$ for an $n$-qubit unitary $U$
can be done
by an Arthur-Merlin protocol with the $F^{-1}$-time Arthur.
However, if we consider the realistic case,
$F\sim2^{-m}$ and $m\gg n$, where $m$ is the number of gates,
the time-complexity of Arthur is $\sim2^m$.
(On the other hand, the exact computation of $|\langle0^n|U|0^n\rangle|^2$
can be done in time $\sim2^n$.)
Moreover, although Ref.~\cite{Google} considered exact sampling of 
$\{p_z'\}_z$,
what a realistic
quantum computer can do is approximately sampling
$\{p_z'\}_z$.
Theorem~\ref{theorem:result2} 
shows that
if we consider the realistic case, $F\sim2^{-m}$ and $m\gg n$,
quantum supremacy does not exist
for approximate sampling of $\{p_z'\}_z$ 
even with the exponentially-small multiplicative
error $\sim 2^{-(m-n)}$.
Theorem~\ref{theorem:result3} 
shows that
if we consider the realistic case, $F\sim2^{-m}$,
quantum supremacy does not exist
for
approximate sampling of
$\{p_z'\}_z$ even with the exponentially-small additive 
error $\sim2^{-m}$.

\section{Discussion}

Our theorems show that if $F\sim2^{-m}$ and $m\gg n$,
or at least $F\sim2^{-m}$,
quantum supremacy does not exist for
approximate sampling of $\{p_z'\}_z$.
In this section, we argue that if $F\sim2^{-m}$ and $m\gg n$,
the standard approach will not work to show
quantum supremacy for exact sampling of $\{p_z'\}_z$.

In the standard proof of quantum 
supremacy~\cite{TD,BS,IQP1,MFF,Kobayashi,KobayashiICALP}, 
we first consider the following
promise problem: given the classical description of an $n$-qubit
$m$-size quantum circuit $U$, and parameters $a$ and $b$, decide
$p_{acc}\ge a$ or $p_{acc}\le b$,
where $p_{acc}$ is the acceptance probability.
In the standard proof of quantum supremacy, we take the promise problem
as the complete problem of a ``strong" quantum class,
such as postBQP, SBQP, or NQP.

We next assume that $p_{acc}'\equiv Fp_{acc}+\frac{1-F}{2^n}$
is exactly classically sampled. It means that there exists a
polynomial-time classical probabilistic algorithm that accepts with
probability $q_{acc}$ such that
$q_{acc}=p_{acc}'$.

If the answer of the promise problem is yes, then
$q_{acc}\ge Fa+\frac{1-F}{2^n}\equiv\alpha$.
If the answer of the promise problem is no, then
$q_{acc}\le Fb+\frac{1-F}{2^n}\equiv\beta$.
In the standard proof of quantum supremacy, we then conclude
that the promise problem is in a ``weaker" class
(such as postBPP, SBP, or NP) that leads to an unlikely
consequence in complexity theory,
such as ${\rm postBQP}\subseteq{\rm postBPP}$,
${\rm SBQP}\subseteq{\rm SBP}$, or
${\rm NQP}\subseteq{\rm NP}$.
However, 
deciding
$q_{acc}\ge \alpha$ or $q_{acc}\le \beta$
seems to be ``more difficult" than the original promise problem:
the original promise problem can be solved in time $\sim2^n$,
while deciding $q_{acc}\ge\alpha$ or $q_{acc}\le\beta$ 
will not be solved in that time
because $\alpha-\beta=F(a-b)=O(2^{-m})$, and $m\gg n$.
Therefore we will not have any unlikely consequence in
this approach.

Although the above argument does not exclude the existence of
a completely new supremacy proof for the exact sampling
of $\{p_z'\}_z$ that works even when $F\sim2^{-m}$ and $m\gg n$,
we can also argue that even if the realistic quantum
computer exactly samples
$\{p_z'\}_z$, it is ``effectively" classically samplable
by the uniform distribution when $F\sim2^{-m}$
unless we can access exponentially many samples. 

To see this,
let us consider the task of distinguishing 
$\rho_0\equiv\frac{I^{\otimes n}}{2^n}$
and 
$\rho_1\equiv F\rho+(1-F)\frac{I^{\otimes n}}{2^n}$,
where $\rho$ is any $n$-qubit state.
Assume that we can measure $k$ copies of $\rho_0$ or $\rho_1$.
Let $\Pi_0$ $(\Pi_1)$ be the POVM element that we conclude
that the actual state is $\rho_0^{\otimes k}$
($\rho_1^{\otimes k}$), where
$\Pi_0+\Pi_1=I^{\otimes nk}$.
The probability $p_{correct}$ that we make the correct decision is
\begin{eqnarray*}
p_{correct}&\equiv&
\frac{1}{2}{\rm Tr}(\Pi_0\rho_0^{\otimes k})
+\frac{1}{2}{\rm Tr}(\Pi_1\rho_1^{\otimes k})\\
&=&
\frac{1}{2}+\frac{1}{2}\Big[
{\rm Tr}(\Pi_0\rho_0^{\otimes k})
-{\rm Tr}(\Pi_0\rho_1^{\otimes k})\Big]\\
&\le&
\frac{1}{2}+
\frac{1}{4}\big\|\rho_0^{\otimes k}
-\rho_1^{\otimes k}\big\|_1\\
&\le&
\frac{1}{2}
+\frac{k}{4}
\big\|\rho_0-\rho_1\big\|_1\\
&=&
\frac{1}{2}+
\frac{k}{4}\Big\|\frac{I^{\otimes n}}{2^n}
-\Big[F\rho+(1-F)\frac{I^{\otimes n}}{2^n}\Big]\Big\|_1\\
&=&
\frac{1}{2}+
\frac{kF}{4}\Big\|
\rho-\frac{I^{\otimes n}}{2^n}\Big\|_1\\
&\le&\frac{1}{2}+\frac{kF}{2}.
\end{eqnarray*}
If $F\sim2^{-m}$ and $k=o(2^m)$, 
$p_{correct}\to \frac{1}{2}$.

\section{Proof of Theorem~\ref{theorem:result}}
Assume that a language $L$ is in BQP. Then
for any polynomial $r$,
there exists a polynomial-time uniformly generated
family $\{V_x\}_x$ of quantum circuits such that
if $x\in L$ then $p_{acc}\ge 1-2^{-r(|x|)}$,
and
if $x\notin L$ then $p_{acc}\le 2^{-r(|x|)}$.
Here
\begin{eqnarray*}
p_{acc}\equiv\langle0^w|V_x|0^w\rangle
\end{eqnarray*}
with $w=poly(|x|)$ is the acceptance probability.

Let $m$ be the number of elementary gates in $V_x$, i.e.,
$V_x=w_mw_{m-1}...w_2w_1$, where each $w_j$ is an elementary gate
(such as $H$, $CNOT$, and $T$, etc.).
Let us consider the following random quantum circuit 
on $n\equiv w+m$ qubits:
\begin{itemize}
\item[1.]
The initial state is $|0^w\rangle\otimes|0^m\rangle$, where we call
the first $w$-qubit register the main register,
and the second $m$-qubit register the ancilla register.
\item[2.]
For each $j=1,2,...,m$, apply $w_j\otimes I$ or 
$\eta_j\otimes X$ with probability
1/2, where $\eta_j$ is any elementary gate,
$w_j$ and $\eta_j$ act on the main register,
and $I$ and $X$ act on the $j$th qubit of the ancilla register.
Thus obtained the final state is
\begin{eqnarray}
\frac{1}{2^m}\sum_{\alpha\in\{0,1\}^m}
\xi_m^{\alpha_m}...\xi_1^{\alpha_1}|0^w\rangle\langle0^w|
(\xi_1^{\alpha_1})^\dagger...(\xi_m^{\alpha_m})^\dagger
\otimes|\alpha\rangle\langle\alpha|,
\label{final}
\end{eqnarray}
where $\alpha\equiv(\alpha_1,...,\alpha_m)\in\{0,1\}^m$ is an $m$-bit string,
$\xi_j^0=w_j$,
and $\xi_j^1=\eta_j$.
\item[3.]
Measure all $n$ qubits in the computational basis.
If all results are 0, accept. Otherwise, reject.
\end{itemize}

If we consider the globally depolarized version,
the state of Eq.~(\ref{final}) is replaced with
\begin{eqnarray*}
\frac{F}{2^m}\sum_{\alpha\in\{0,1\}^m}
\xi_m^{\alpha_m}...\xi_1^{\alpha_1}|0^w\rangle\langle0^w|
(\xi_1^{\alpha_1})^\dagger...(\xi_m^{\alpha_m})^\dagger
\otimes|\alpha\rangle\langle\alpha|
+(1-F)\frac{I^{\otimes n}}{2^n}.
\end{eqnarray*}
The acceptance probability 
$p_{acc}'$ 
is 
\begin{eqnarray*}
p_{acc}'=\frac{Fp_{acc}^2}{2^m}+\frac{1-F}{2^n}.
\end{eqnarray*}

Assume that there exists a classical $poly(n)$-time probabilistic algorithm
that accepts with probability $q_{acc}$ such that 
$|p_{acc}'-q_{acc}|\le \epsilon p_{acc}'$,
where $\epsilon<1$ is a constant.
Then, if $x\in L$,
\begin{eqnarray*}
q_{acc}&\ge&(1-\epsilon)p_{acc}'\\
&=&(1-\epsilon)\Big(\frac{Fp_{acc}^2}{2^m}+\frac{1-F}{2^n}\Big)\\
&\ge&(1-\epsilon)F2^{-m}(1-2^{-r})^2,
\end{eqnarray*}
and if $x\notin L$, 
\begin{eqnarray*}
q_{acc}&\le&(1+\epsilon)p_{acc}'\\
&=&(1+\epsilon)\Big(\frac{Fp_{acc}^2}{2^m}+\frac{1-F}{2^n}\Big)\\
&\le&(1+\epsilon)\Big(F2^{-2r-m}+\frac{1-F}{2^n}\Big)\\
&=&2^{-m}(1+\epsilon)F\Big(2^{-2r}+\frac{1-F}{F2^w}\Big).
\end{eqnarray*}
If $r$ and $w$ are sufficiently large, $L$ is in SBP.
\fbox

Note that although here we have considered constant $F$,
the same result also holds for other ``not so small" $F$
such as $F=\frac{1}{poly(m)}$.

\section{Proof of Theorem~\ref{theorem:result2}}
Let us take $\epsilon=F2^{n+2}$.
For any $z\in\{0,1\}^n$,
\begin{eqnarray*}
\Big|p_z'-\frac{1}{2^n}\Big|
=\Big|\Big(Fp_z+\frac{1-F}{2^n}\Big)
-\frac{1}{2^n}\Big|\le F\Big(1+\frac{1}{2^n}\Big)< \epsilon p_z',
\end{eqnarray*}
where in the last inequality, we have used
\begin{eqnarray*}
\epsilon p_z'-F\Big(1+\frac{1}{2^n}\Big)
&=&
\epsilon\Big(Fp_z+\frac{1-F}{2^n}\Big)-F\Big(1+\frac{1}{2^n}\Big)\\
&\ge&\frac{\epsilon(1-F)}{2^n}-F\Big(1+\frac{1}{2^n}\Big)\\
&=&\frac{F2^{n+2}(1-F)}{2^n}-F\Big(1+\frac{1}{2^n}\Big)\\
&=&4F(1-F)-F\Big(1+\frac{1}{2^n}\Big)\\
&>0&.
\end{eqnarray*}
\fbox

\section{Proof of Theorem~\ref{theorem:result3}}
\begin{eqnarray*}
\sum_{z\in\{0,1\}^n}\Big|p_z'-\frac{1}{2^n}\Big|
= F\sum_{z\in\{0,1\}^n}\Big|p_z-\frac{1}{2^n}\Big|
\le 2F.
\end{eqnarray*}
\fbox

\acknowledgements
TM is supported by 
MEXT Quantum Leap Flagship Program (MEXT Q-LEAP) Grant Number JPMXS0118067394,
JST PRESTO No.JPMJPR176A,
and the Grant-in-Aid for Young Scientists (B) No.JP17K12637 of JSPS. 
YT is supported by
MEXT Quantum Leap Flagship Program (MEXT Q-LEAP) Grant Number JPMXS0118067394.


\begin{thebibliography}{99}

\bibitem{TD}
B. M. Terhal and D. P. DiVincenzo,
Adaptive quantum computation, constant depth
quantum circuits and Arthur-Merlin games.
Quant. Inf. Comput. {\bf4}, 134 (2004).

\bibitem{BS}
S. Aaronson and A. Arkhipov, The computational complexity
of linear optics. Theory of Computing {\bf9}, 143 (2013).

\bibitem{IQP1}
M. J. Bremner, R. Jozsa, and D. J. Shepherd,
Classical simulation of commuting quantum computations
implies collapse of the polynomial hierarchy.
Proc. R. Soc. A {\bf467}, 459 (2011).

\bibitem{IQP2}
M. J. Bremner, A. Montanaro, and D. J. Shepherd,
Average-case complexity versus approximate
simulation of commuting quantum computations.
Phys. Rev. Lett. {\bf117}, 080501 (2016).

\bibitem{KL}
E. Knill and R. Laflamme,
Power of one bit of quantum information.
Phys. Rev. Lett. {\bf81}, 5672 (1998).

\bibitem{MFF}
T. Morimae, K. Fujii, and J. F. Fitzsimons,
Hardness of classically simulating the one clean qubit model.
Phys. Rev. Lett. {\bf112}, 130502 (2014).

\bibitem{M}
T. Morimae, Hardness of classically sampling one
clean qubit model with constant total variation distance
error.
Phys. Rev. A {\bf96}, 040302(R) (2017).

\bibitem{Kobayashi}
K. Fujii, H. Kobayashi, T. Morimae, H. Nishimura,
S. Tamate, and S. Tani,
Impossibility of classically simulating one-clean-qubit
model with multiplicative error.
Phys. Rev. Lett. {\bf120}, 200502 (2018).

\bibitem{KobayashiICALP}
K. Fujii, H. Kobayashi, T. Morimae, H. Nishimura, S. Tamate,
and S. Tani,
Power of quantum computation with few clean qubits.
Proceedings of 43rd International
Colloquium on Automata, Languages,
and Programming (ICALP 2016),
pp.13:1-13:14 (2016).

\bibitem{random1}
A. Bouland, B. Fefferman, C. Nirkhe, and U. Vazirani,
Quantum supremacy and the complexity of random circuit
sampling.
Nature Phys. {\bf15}, 159 (2019).

\bibitem{random2}
R. Movassagh,
Cayley path and quantum computational supremacy:
a proof of average-case $\#$P-hardness of random circuit sampling
with quantified robustness.
arXiv:1909.06210

\bibitem{Dalzell}
A. M. Dalzell, A. W. Harrow, D. E. Koh, and R. L. La Placa,
How many qubits are needed for quantum computational supremacy?
arXiv:1805.05224

\bibitem{DalzellPhD}
A. M. Dalzell,
Bachelor thesis, MIT (2017).

\bibitem{Huang}
C. Huang, M. Newman, and M. Szegedy,
Explicit lower bounds on strong quantum simulation.
arXiv:1804.10368

\bibitem{Huang2}
C. Huang, M. Newman, and M. Szegedy,
Explicit lower bounds on strong simulation of quantum circuits
in terms of $T$-gate count.
arXiv:1902.04764

\bibitem{MorimaeTamaki}
T. Morimae and S. Tamaki,
Fine-graiend quantum computational supremacy.
Quant. Inf. Comput. {\bf19}, 1089 (2019).

\bibitem{Google}
F. Arute et al. 
Quantum supremacy using a programmable superconducting processor.
Nature {\bf 574}, 505 (2019).




\bibitem{SBP}
E. B\"{o}hler, C. Gla\ss er, and D. Meister,
Error-bounded probabilistic computations between MA and AM.
J. Comput. Syst. Sci. {\bf72}, 1043 (2006).

\bibitem{Raz}
R. Raz and A. Tal,
Oracle separation of BQP and PH.
Proceedings of the 51st Annual ACM SIGACT Symposium on Theory
of Computing (STOC 2019), pp.13-23 (2019).

\bibitem{nonclean}
T. Morimae, K. Fujii, and H. Nishimura,
Power of one nonclean qubit.
Phys. Rev. A {\bf95}, 042336 (2017).

\end{thebibliography}
\end{document}